\documentstyle[aps, prl, twocolumn, floats, epsfig]{revtex}

\title{    Transport Properties of One-Dimensional Hubbard Models}
\author{   S. Kirchner$^{1,2}$, H.G. Evertz$^{1,3}$, and W. Hanke$^1$}
\address{  $^1$\ Inst.\ f.\ Theoretische Physik, Univ.\ W\"urzburg, 97074 W\"urzburg, Germany\\
           $^2$\ Inst.\ f.\ Theorie d.\  Kondensierten Materie, Univ.\ Karlsruhe,  76128 Karlsruhe,  Germany,\\
           $^3$\ ISSP, Univ.\ of Tokyo, Roppongi 7-22-1, Minato-ku, Tokyo 106, Japan}
\date{April 15, 1998}

\newcommand{\note}[1]{{}}
\newcommand{\beq}[1]{\begin{equation}\label{#1}}
\newcommand{\beqn}[1]{\begin{eqnarray}\label{#1}}
\newcommand{\Label}[1]{\label{#1}}
\newcommand{\eeq}{\end{equation}}
\newcommand{\eeqn}{\end{eqnarray}}
\newcommand{\Eq}[1]{Eq.\ (\ref{#1})}
\newcommand{\eq}[1]{eq.\ (\ref{#1})}
\newcommand{\Fig}[1]{Fig.\ \ref{#1}}
\newcommand{\fig}[1]{fig.\ \ref{#1}}
\newcommand{\mod}{\mbox{ mod }}
\newcommand{\bigsum}[1]{{\mbox{$\displaystyle\sum_{\scriptsize #1}$}}}
\newcommand{\mystack}[2]{{\hskip-2ex\begin{array}{c}{#1}\\{#2}\end{array}\hskip-2ex}}
\def\tr{\mathop{\rm tr}}

\def\Im{\mathop{\rm Im}}
\def\Re{\mathop{\rm Re}}
\def\abs#1{\left| #1\right|}
\def\bra#1{\left\langle #1\right|}
\def\ket#1{\left| #1\right\rangle}
\def\VEV#1{\left\langle #1\right\rangle}

\begin{document}

\twocolumn[\hsize\textwidth\columnwidth\hsize\csname @twocolumnfalse\endcsname
\maketitle
\begin{abstract}
We present results for the zero and finite temperature Drude weight $D(T)$ 
and for the Meissner fraction
of the attractive and the repulsive Hubbard model,
as well as for the model with next nearest neighbor repulsion.
They are based on Quantum Monte Carlo studies and on the Bethe ansatz.
We show that the Drude weight is well
defined as an extrapolation on the imaginary frequency axis,
even for finite temperature.
The temperature, filling, and system size dependence of $D$ is obtained.
We find counterexamples to a conjectured
connection of dissipationless transport 
and integrability of lattice models.
\end{abstract}
]

\section{Introduction}
An ideal conductor is characterized at zero temperature
by a nonvanishing Drude weight $D$ in the real part of the conductivity,
\beq{cond}
  \Re \{\sigma(\omega)\}\,=\, D \, \delta(\omega) \,+\, \sigma_{reg}(\omega) \,,
\eeq
as first introduced by Kohn in the context of the Mott transition \cite{Kohn64}.
For finite temperature, the Drude weight $D(T)$ can be introduced by a
formal extension of eq. (\ref{cond}): 
\beq{ext}
  \Re \{\sigma(\omega,T)\}\,=\, D(T) \, \delta(\omega) \,+\, \sigma_{reg}(\omega,T).
\eeq
A superconductor is characterized by an additional
quantity probing the Meissner effect, 
the superfluid density \cite{Scalapino-White-Zhang93}.
There are similar transport quantities for spin degrees of freedom \cite{Kopietz97}.

Recently there has been a lot of interest in transport properties
at finite temperature, and especially in the Drude weight,
since the High-T$_{c}$ materials exhibit an unusual
low-frequency behavior of the optical conductivity in the underdoped
regime \cite{Thomas91}. 
Based on analytical and numerical results, 
a very interesting 
possible connection between the integrability of a lattice-model and its
finite-temperature Drude weight has been proposed, conjecturing 
\cite{Castella-Zotos-Prel95,Zotos-Prelovsek96,Zotos-Castella96,Zotos-Naef-Prelovsek97}
that an integrable system is characterized by a finite Drude
weight $D(T> 0)\neq 0$ when $D(T=0) \neq 0$,
and remains an ideal insulator $D(T>0)=0$ when $D(T=0)=0$,
whereas a nonintegrable system should exhibit a vanishing Drude weight at $T>0$.

The results of our finite temperature Quantum Monte Carlo simulations
do not confirm such a connection in the repulsive Hubbard model with and
without next nearest neighbor interaction.
We also show that the Drude weight can be extracted directly 
by an extrapolation of the current-current correlations on Matsubara
frequencies, even at finite temperature, thus avoiding 
an analytic continuation to real frequencies.

We investigate the one-dimensional Hubbard mo\-del on a ring of $L$
sites threaded by a flux $L \Phi$. The Hamiltonian is
\beqn{hubb-model}
    H_H(\Phi ) \,&=&\, -t \sum_{i, \sigma} (c_{i,\sigma}^{\dagger}\,c_{i+1,
 \sigma}^{ }e^{-ie\, \Phi(\vec{x}_{i} ,t)}\\
 \,&+&\,  c_{i+1,\sigma}^{\dagger} \, c_{i,\sigma}^{ }e^{ie\, \Phi(\vec{x}_{i},t)})  
 \,+\, U \, \sum_{i} {n}_{i \uparrow} n_{i \downarrow} \nonumber,
\eeqn
where $c,c^\dagger$ are annihilation and creation operators
and the Peierls phase $\Phi(\vec{x}_{i},\omega)=\int_{\vec{x}_{i}}^{\vec{x}_{i+1}}\,
\vec{A}(\vec{z},\omega)\,d \vec{z}$ in general is a function of position and frequency.
We set $\hbar$, $c$ and the lattice spacing to unity, and we 
specify energies  in units of $t$.
We use periodic boundary conditions $c_{i+L,\sigma}=c_{i,\sigma}$.

The Drude weight of the one-dimensional Hubbard model
at zero temperature has been investigated in several papers,
including studies of the scaling behavior of $D$ 
at half-filling by Stafford, Millis, and Shastry \cite{Stafford-Millis-Shastry90},
close to half-filling by Stafford and Millis \cite{Stafford-Millis93}
and by Fye et al.\ \cite{Fye-Martins-Hanke91}. 
For arbitrary filling, results were given by Schulz \cite{Schulz90},
and by Fye et al. with the Lanczos method for small systems \cite{Fye-Martins-Hanke91}. 
R\"omer and Punnoose \cite{Roemer95} computed both Drude weight and spin stiffness.
Some related properties of charge and spin currents at finite temperature were
recently computed by Peres et al.\ \cite{Peres97} in a perturbation theory
based on the Bethe ansatz.
For the limit $L\rightarrow\infty$, expressions for the Drude weight based on 
the Bethe ansatz at finite temperature were very recently given by 
Fujimoto and Kawakami \cite{Fujimoto97}.

In section \ref{DRUDE} we discuss representations of the finite
temperature Drude weight,
and show that it can be obtained by an extrapolation purely in 
imaginary frequencies.
In section \ref{INT} we briefly review the conjectured connection to integrability.
In section \ref{BETHE} we compute the filling dependence of $D(T=0)$ at zero temperature 
via the Bethe ansatz equations, and use these equations to give an approximation 
to $D(T)$ at half-filling for low temperature and small system sizes.
In section \ref{QM} we present a systematic Quantum Monte Carlo study of the
finite temperature Drude weight in the Hubbard model, both repulsive
and attractive. Results for a closely
related property, the Meissner fraction,
are presented in section \ref{MF},
and we study $D(T)$ in the extended Hubbard model in section \ref{HEXT}.

\section{\label{DRUDE} The Drude Weight}

Applying a Kramers-Kronig relation to \eq{ext} results in
\beq{alt}
  D(T) \,=\, \pi \, \lim_{\omega \rightarrow 0} (\omega \Im\{\sigma(\omega,T)\}).
\eeq
Using linear response, the Drude weight is given by \cite{Scalapino-White-Zhang93}
\beq{drudelinres}
   \frac{D(T)}{\pi e^2} \,=\, -\VEV{k_{x}} \,-\, \Re\{ \Lambda(0, \omega \rightarrow 0)\}
\eeq
at $\Phi =0$, where $\VEV{k_{x}}$ denotes the average kinetic energy per site and
$\Lambda(\vec{q} ,\omega)$ is the current-current correlation
function in frequency space \cite{Scalapino-White-Zhang93} (see appendix A
for  details). 
\\
There are two different ways to obtain the Drude weight from the imaginary frequency
correlation function.
One can perform an analytic continuation of the data to real frequency 
with the Maximum Entropy method to obtain $\Im\Lambda(\omega)/\omega$,
and then use the f-sum rule:
\beq{fsum}
  \frac{D}{\pi e^2} = -\VEV{k_x}  
                      -\frac{2}{\pi} \int_{0+}^\infty d\omega \frac{\Im\Lambda(\omega)}{\omega}.
\eeq

Alternatively, one can work entirely on the imaginary frequency axis:
The analytical continuation of the
current-current correlation function $\Lambda$,
given in \Eq{ccmats}, is valid
in the continuous upper plane,
including the imaginary axis at frequencies different from the Matsubara frequencies.
One can therefore take the limit $\omega\rightarrow 0$ for $\Lambda$ either along
the real axis, or purely on the imaginary axis, even at finite temperature.
The latter version eliminates the need for an analytic continuation
(e.g.\ via Maximum Entropy) from data on the imaginary Matsubara frequencies
onto real frequencies.
In the present paper we employed this procedure.
We have verified that it produces results compatible with those using the f-sum rule, 
but with smaller errors.

The generalization of $D$ to finite temperatures can also be
achieved by defining \cite{Zotos-Castella96} 
\begin{equation}
  \label{ext2}
  D(T) \, = \, \frac{\pi}{L} \sum_{n} \frac{e^{-\beta E_{n}}}{Z}\,
    \frac{\partial^{2}\,E_{n}(\Phi)}{\partial\,\Phi^{2}}\Big|_{\Phi=0,\vec{q}=0},
\eeq
where $Z=\tr e^{-\beta H}$ is the partition function. 
In the limit $T\rightarrow 0$, this immediately reduces to Kohn's
Drude weight 
\beq{Kohn}
D(T=0)=\frac{\pi}{L}\frac{\partial^{2}\,E_{0}(\Phi)}{\partial\,\Phi^{2}}
                                           \Big|_{\Phi=0,\vec{q}=0}.
\eeq
The equivalence of 
\eq{drudelinres} and \eq{ext2} within perturbation theory is shown in appendix A.

\section{\label{INT} Connection to integrability} 
A model is usually called integrable if the energy eigenvalues are distributed
according to a Poisson distribution. For a non-integrable model, the eigenvalues follow a
GOE distribution (Gaussian Orthogonal Ensemble)
\cite{Poilblanc-Ziman-Mila93}. 
It can be shown  for lattice models which are solvable by the
Bethe ansatz that the eigenvalues are Poisson distributed \cite{Berry-Tabor77}. 
Hence the Hubbard model \eq{hubb-model} is integrable.
In section \ref{HEXT} we will study the extended Hubbard-model with nearest 
neighbor repulsion, which is non-integrable \cite{Montambaux-Poilblanc93}.

Based on analytical and numerical results, Zotos et al.\
\cite{Castella-Zotos-Prel95,Zotos-Prelovsek96,Zotos-Castella96,Zotos-Naef-Prelovsek97}
have conjectured a very interesting
connection between the integrability of a lattice model and its
finite-temperature Drude weight,
stating that for a one-dimensional model
the finite temperature Drude weight in the thermodynamic limit is
\cite{Zotos-Prelovsek96,Zotos-Castella96}
\begin{itemize}
 \item[(1)] nonzero for an integrable system when it is nonzero at $T=0$,
 \item[(2)] zero for an integrable system when it is zero at $T=0$, and
 \item[(3)] zero for a non-integrable system.
\end{itemize}
Conjecture (2) is different from the original 
suggestion \cite{Castella-Zotos-Prel95}
of a direct equivalence between integrability and finite Drude weight
at $T>0$ in one-dimensional systems.
It was motivated by an explicit example 
in which the authors computed the Drude
weight for a one-dimensional model of interacting fermions 
in the insulating regime with a ``Mott-Hubbard'' gap,
and
found a vanishing Drude value in the
insulating regime even at high temperatures.

Note that there is no rigorous proof of a connection between 
integrability and a finite Drude weight.
However, subject to some assumptions, 
Zotos et al.\ have shown in their most recent paper \cite{Zotos-Naef-Prelovsek97} 
that the Drude weight is finite whenever there exists an operator $A$
such that $[A,\, H]=0$ and
\beq{schwarzrule}
  \lim_{L\rightarrow\infty} \frac{\beta}{L} \frac{\VEV{j\,A}^{2}}{\VEV{A^{2}}} \ne 0.
\eeq
Thus, a weaker condition than integrability might suffice to make the 
Drude weight finite.

\section{\label{BETHE} Zero and low temperature:  Bethe ansatz}
%
We employ the Bethe ansatz equations for the Hubbard model,
first obtained by Lieb und Wu in 1968 \cite{Lieb-Wu68}, to calculate 
$D(T)$ and its filling dependence exactly at zero temperature,
and to provide an approximation for $D(T)$ at low finite temperature.
As was shown by Shastry and Sutherland\cite{Shastry-Sutherland90}, 
the Hubbard model with twisted periodic boundary conditions
\beq{twistbdncond}
    \Psi(\ldots, \vec{r}+L \vec{x}, \ldots) \,=\, \exp[i\phi] \, 
    \Psi(\ldots,\vec{r}, \ldots)
 \eeq
can be solved with an appropriate ansatz for the wave function.
Here $\phi=e L\, \Phi$ is the overall phase aquired along the chain 
of length $L$ \cite{Remark}.
The Bethe ansatz equations derived by Shastry and Sutherland are:
\beq{bethegl1}
    Lk_{n}\,=\, 2\pi I_{n} \,+\, \phi \,+\, 2\, \sum_{j=1}^{M} 
\arctan[4(\Lambda_{j} - \sin k_{n})/U]
\eeq
\beqn{bethgl2}
  2 \sum_{n=1}^{N}\, \arctan
&&[4(\Lambda_{j} - \sin k_{n})/U] \\  
&=& \, 2\pi J_{j} \,+ \, 2 \sum_{i \neq j}^{M}
\arctan[2(\Lambda_{j}
       -\Lambda_{i})/U], \nonumber
\eeqn
where N is the total number of electrons, 
$M$ is the number of electrons with spin down, 
the $\Lambda_{j}$ are parameters associated with the spin dynamics, 
and the quantum numbers 
$I_{n}$ and $J_{m}$ characterize charge and spin excitations. 
The quantum numbers have to be chosen such 
that \cite{Stafford-Millis93,Andrei92} 
\begin{itemize}
\item[]  $I_{n}$ is integer$/$half-integer if $M$ is even$/$odd, 
\item[]  $J_{m}$ is integer$/$half-integer if $N-M$ is odd$/$even,  
\end{itemize}
all $I_n \, (n=1,..,N)$   (all $J_m \,(m=1,..,M)$) are different from each other,
and
\beqn{IJ}
         |I_n|\le (L-1)/2 \\
         |J_m|\le (N-M)/2 \,. \nonumber
\eeqn
The energy of a state with given quantum numbers is given by \cite{Andrei92} 
\beq{energy}
  E\,=\,-2\, \sum_{n} \cos(k_{n}) \,.
\eeq

Following ref.\ \cite{Stafford-Millis-Shastry90}, we then calculate the Drude weight
{}from \eq{ext2} as 
\beq{Drudeweight}
    \, D \,=\, 2\pi e^{2} L \sum_{n} \left \{ \cos k_{n} \left [\frac{dk_{n}}{d \phi}
 \right ]^{2 } + \sin k_{n} \frac{d^{2} k_{n}}
   {d \phi^{2}} \right \} \Bigg{|}_{\phi=0}.
\eeq

\subsection{Zero Temperature}
At {\em half-filling and zero temperature} the asymptotic $L$-dependence of the Drude-weight is
\beq{drudskal}
  D(L;T=0) \,=\, -(-1)^{L/2} C(U)L^{1/2} \exp[-L/ \zeta(U)],
\eeq
(L even) as was shown by Stafford and Millis \cite{Stafford-Millis93}.
The correlation length
$\zeta$ is defined by \eq{drudskal}.
For ${L \rightarrow \infty}$ the Drude weight in \eq{drudskal}
vanishes, indicating the finite charge gap at half-filling.
We show results for half-filling in fig.\ \ref{Abb2.1}.
For $L \mod 4 = 0$  the ring becomes paramagnetic.
This has already been discussed 
in refs.\ \cite{Stafford-Millis-Shastry90} and \cite{Fye-Martins-Hanke91}.
We find that the effective correlation length $\zeta\approx 4.25$ 
at the system sizes of \fig{Abb2.1} is slightly  larger than the 
value $\zeta=4.06$ at $U=4$ found numerically in \cite{Stafford-Millis-Shastry90}.

%
\begin{figure}[t]
\psfig{file=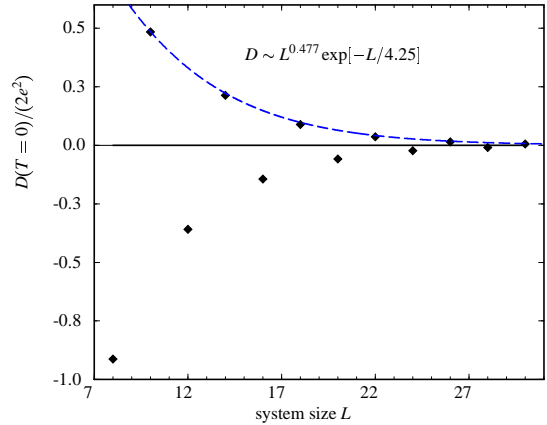,width=8.5 cm}
\caption{\Label{Abb2.1} 
Scaling dependence of $D(T=0)$ at $U=4$ and half-filling.
}
\end{figure}

{\em Away from half-filling} the Hubbard model describes an ideal metal,
with a finite Drude weight at $L\rightarrow\infty$.
We show the system size dependence at quarter filling in \fig{fig2a}.
We find that the L-dependence is very similar to the half-filled 
case, shifted by $D(L=\infty)$.
Even though \eq{drudskal} was only derived for the half-filled case,
it fits the approach to $D(L=\infty)$ very well,
albeit with a different exponent for $L$ and a very large value of $\zeta$,
as shown in the inset of \fig{fig2a}.
%
\begin{figure}[t]
  \psfig{file=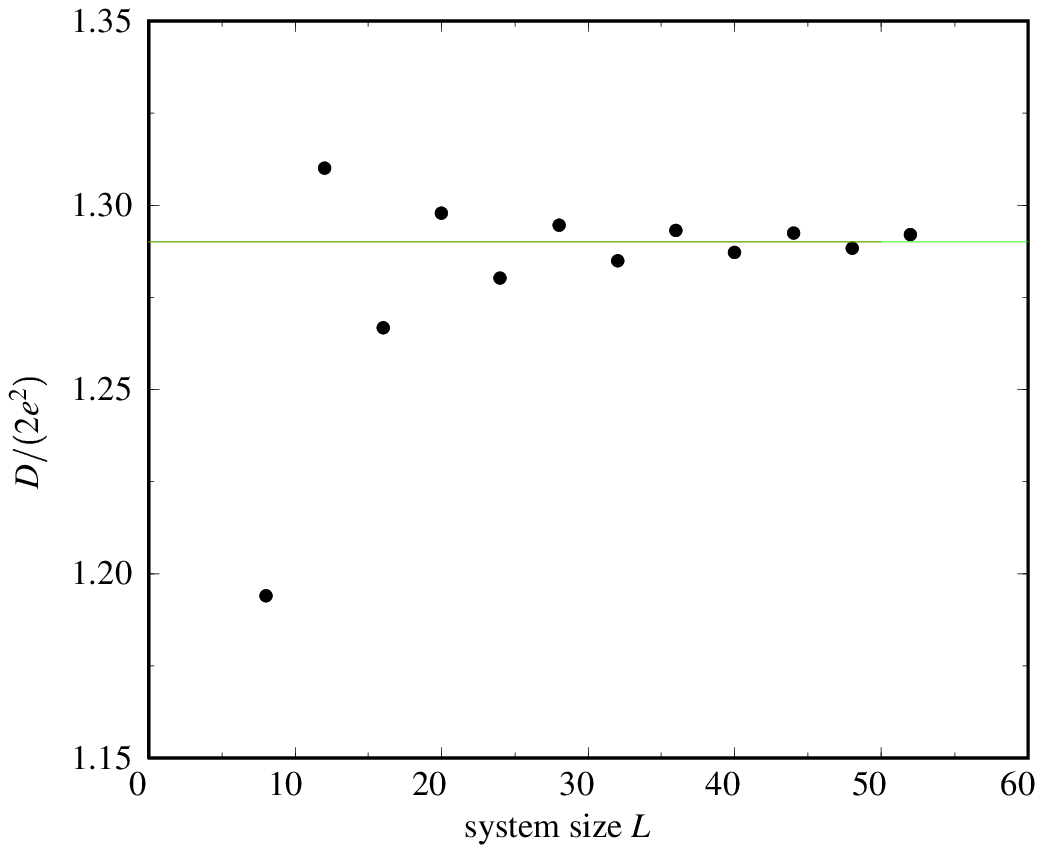,width=8 cm}
  \vskip-3.7cm\hskip3cm\psfig{file=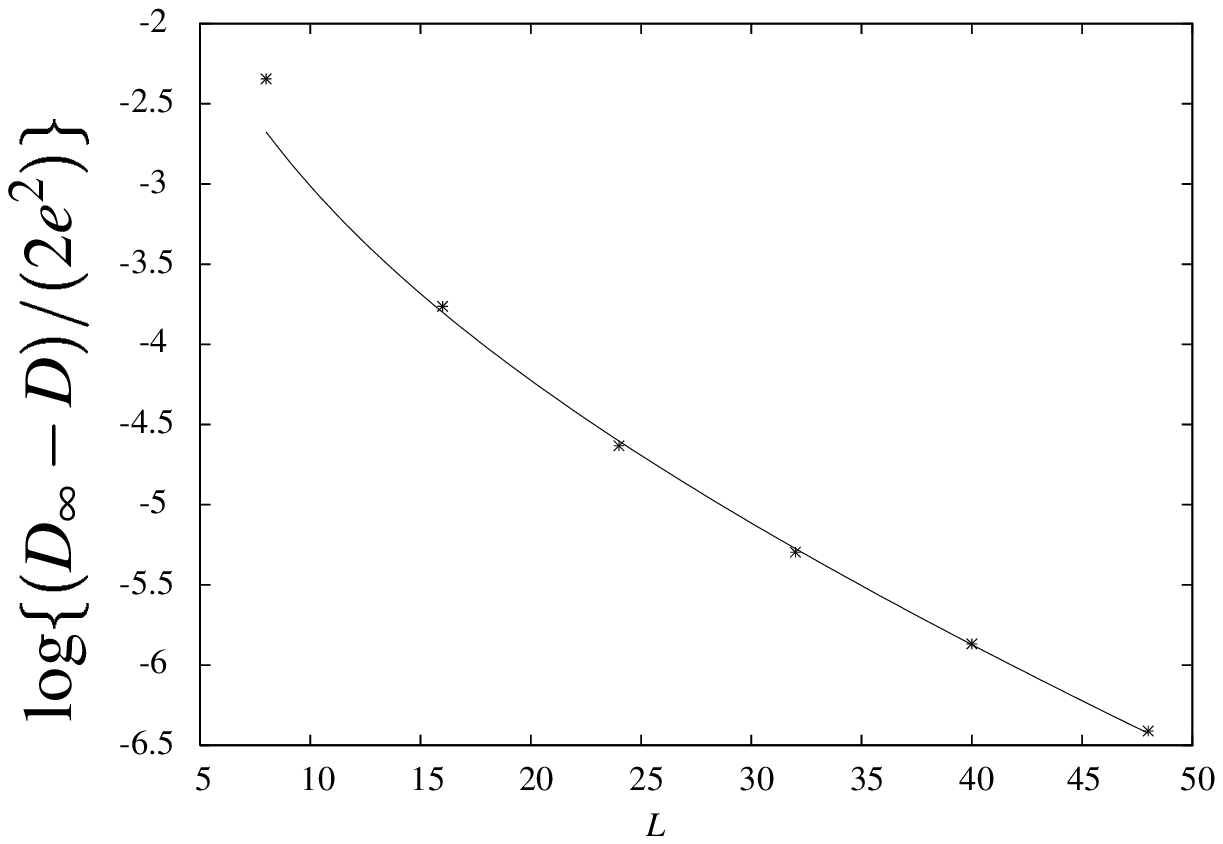,height=2.6cm}
  \vskip.9cm
\caption{\Label{fig2a}
  The Drude weight at $U\!=\!4$ and quarter filling $\nu\!=\!\frac{1}{2}$
  scales towards a finite value $D(L\!=\!\infty)/$$(2e^2)=1.29$.
  The inset shows the scaling dependence 
  $(D(L=\infty)-D(L))/(2e^2)$ $= L^{-1.12(1)} \exp[-L\cdot 0.04(1)] $ 
  on a logarithmic scale (at $L \mod 4 = 0$).
}
\end{figure}

We determined the complete filling dependence of the Drude weight
by solving the Bethe ansatz equations numerically 
for system sizes up to  $L=80$,
and extrapolating to infinite system size using \eq{drudskal}.
The results are shown in \fig{figtwo}.
As expected, the Drude weight vanishes for $\nu=0$ (no electrons) 
and $\nu=1$ ($L$ electrons).
For small system sizes up to $L=12$ 
the filling dependence has already been reported by
Fye et al.\ using Lanczos techniques \cite{Fye-Martins-Hanke91}.
Solving the Bethe ansatz equations enables us to access much larger
systems  (up to $L=80$) and 
thus to obtain a better  extrapolation to the thermodynamic limit.
The Drude weight for fixed system size $L=100$ has also
been obtained by  R\"omer and Punnoose \cite{Roemer95}, and our results agree.
%
\begin{figure}[t]
\psfig{file=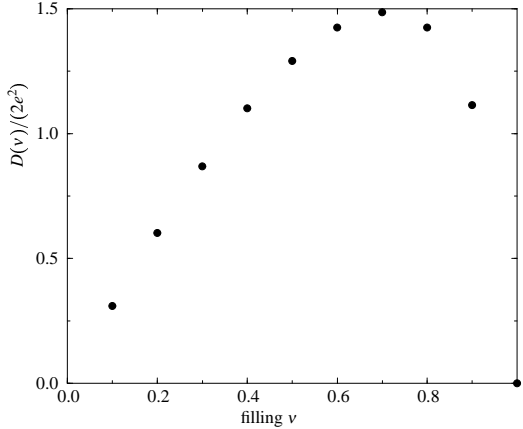,width=8.5cm}
\caption{\Label{figtwo} 
The filling dependence of $D$ at $U=4$ and $T=0$,
extrapolated to infinite system size.
}\end{figure}

\subsection{Partial Drude weight at finite temperature}
We now turn to finite temperatures and half-filling.
Whereas ground state properties depending only on the energy can be calculated easily
using the Bethe ansatz equation \eq{bethegl1},
dynamical properties are much harder to determine 
via the usual Bethe ansatz.
One recent method by Kl\"{u}mper et al.\  reduces the problem to finding
the solution of two coupled nonlinear integral equations
\cite{Kluemper-Suzuki97}. 
Here we will be content with a much simpler procedure to calculate 
the spin triplet contribution to the  finite temperature Drude weight at half-filling.
We neglect excitations across the charge gap.
As we will see, this approximation is valid for small values of
$LT/U$, i.e.\ for low temperatures and not too large systems.

Low energy excitations are described by variations of the quantum numbers
around the ground state configuration.
We choose $N \mod 4 = 2$ whence the ground state is a 
singlet\cite{Stafford-Millis93}. 
As excitations around this state we consider spin triplet states, with $M=N/2-1$,
that is, we consider the $S=1$ and $S_{z}=-1$ excitations. To include singlet excitations
one has to refer to the general Bethe ansatz equations introduced by Takahashi \cite{Takahashi72},
with a phase included as in ref.\ \cite{Peres97}.
The generalized equations describing all four branches of singlet and triplet excitations
(the afore mentioned ones and the $S=1$ and $S_{z}=0$ excitations) are given in
\cite{Carmelo-Peres97}.\\
Due to the charge gap at half-filling, there are no low lying charge exitations. 
The charge quantum numbers $I_n$ therefore remain unchanged from the ground state.
For the spin quantum numbers we get from \eq{IJ} with $M=N/2-$1 that
\beq{maxexcit}
  |J_{j}|  \leq \frac{N}{4} \,,
\eeq
and  the $J_j$ are now half-integer.
Thus there are $N/2+1$ possible values for the $M=N/2-1$ spin quantum numbers.
All possibilities to
distribute the $J_{j}$'s according to \eq{maxexcit} that differ
{}from the ground state constellation describe excited states. 
These are the lowest possible excitations. 
%
In addition, there is a singlet excitation, degenerate in energy with the three
triplet excitations\cite{Andrei92}.
This singlet excitation also contains two holes in the spin
quantum numbers compared with the ground state constellation. It is described
by complex quantum numbers.
Due to the degeneracy, its contribution 
just multiplies the partition function by a constant factor.

The quantity $D^{triplet}(\beta,L)$ calculated via \eq{Drudeweight} using only the
triplet spin excitations is of course only part of the Drude weight, but 
as we will see  it should be a good approximation for small temperatures
and small system sizes.

For the temperature and system size dependence of $D^{triplet}(\beta,L)$ (at $L\mod 4=2$)
we find again a relation similar to the $T=0$ result 
of Stafford and Millis \cite{Stafford-Millis93}, \eq{drudskal}:
\beq{Dspin}
  \frac{D^{triplet}(\beta,L)}{2e^2} \,=\, a(\beta)\,L^{b}\,\exp[-c\cdot L],
\eeq
where only $a(\beta)$ is a function of the temperature.
The results for $D^{triplet}$ are shown in  \fig{three}, and the fitted parameters
$a$ and $c$ in table \ref{tab:1}.

\begin{figure}[t]
 \psfig{file=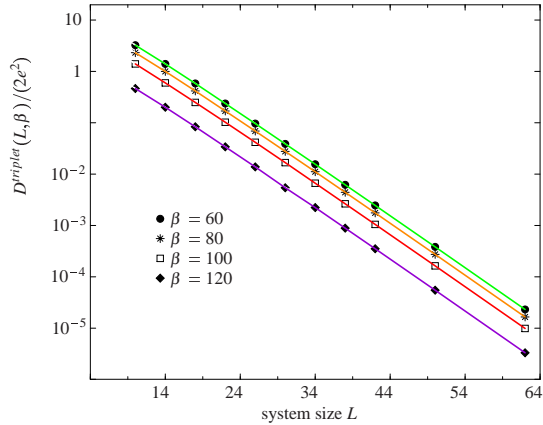,width=8.5 cm}
\caption{\Label{three} 
 Scaling behavior of  $D^{triplet}(T)$ as a function of $L$
 for finite temperatures and $U=4$, $\nu=1$.
}\end{figure}
\begin{table}
    \begin{tabular}[c]{|c||c|c|}
 \hline
 Inverse Temperature  & $a$ & $c$  \\ \hline
 $\beta= 60$ & $16.65 \pm 0.16 $ & $0.2400 \pm 0.0003\;$ \\
 $\beta= 80$ & $11.90 \pm 0.11 $ & $0.2400 \pm 0.0003\;$ \\
 $\beta=100$ & $7.140 \pm 0.068$ & $0.2400 \pm 0.0003\;$ \\
 $\beta=120$ & $2.379 \pm 0.023$ & $0.2400 \pm 0.0003\;$ \\ \hline
    \end{tabular}\vskip2mm
 \caption{ \Label{tab:1}  Parameters $a$ and $c$ 
      obtained by fitting the curves in \fig{three} to \eq{Dspin}.
}\end{table}

We conclude that in the thermodynamic limit, $L \rightarrow \infty$, triplet
excitations do not contribute to the Drude weight.
This also remains true for higher spin excitations \cite{Ref}.

For the $\beta$-range studied, the exponent 
$b=0.336\pm0.005$ and the length scale $c^{-1}=4.17 \pm 0.01$
are independent of temperature. They differ somewhat in value from the $T=0$
result for the complete Drude weight $D(T=0)$ (see \eq{drudskal} and \fig{Abb2.1}).
The amplitude $a(\beta)$ of $D^{triplet}$ is precisely linear in $\beta$
for the $\beta$-range studied.
\beq{abeta}
  a(\beta)=(30.911 \pm 0.007)-(0.23771\pm 0.00009\,)\cdot \beta.
\eeq
This is surprising, since it resembles the high temperature behavior 
generally expected of the complete Drude weight \cite{Castella-Zotos-Prel95}.

We found that the interaction dependence of $D^{triplet}$ for a given temperature and
system size can be approximated by
\beq{Udependence}
  D^{triplet}(U)\,\sim\, U^{-d} \cdot \exp[-f \cdot U],
\eeq
where the parameters $d$ and $f$ depend on system size.

Recently we received a paper by Fujimoto et al.\cite{Fujimoto97}
with exact expressions for the Drude weight at finite temperature
in the limit $L\rightarrow\infty$, based on the Bethe ansatz.
The Drude weight is finite at finite temperature, in agreement with our results.
The authors compute explicitely the leading contributions to D($L=\infty$,T)
at small temperatures,
which at half filling are, as expected, proportional
to \ $\exp[-\Delta_{MH} \beta]$, where $\Delta_{MH}$ is the Mott-Hubbard gap.

Since the quantity $D^{triplet}(\beta,L)$ vanishes exponentially as $L \rightarrow \infty$,
the spin contributions to the Drude weight 
considered here should dominate over the charge excitations
for mesoscopic systems with 
$\beta\gg 1$ (so that higher excitations are suppressed) and
\beq{dom}
  cL \ll \beta U \,
\eeq
i.e.\ for low temperatures $T$ and small values of $LT/U$.

\section{\label{QM} Finite Temperature: QMC results}
In this section we present our results for the Drude weight at large finite temperature 
and compare them with the prediction for the temperature and system size
dependence obtained in the previous section.

The simulations were carried out using the grand canonical Quantum
Monte-Carlo method \cite{Hirsch83}. 
The calculation yields the
current-current correlation function at discrete imaginary times
$\tau_{i}$. Performing a fourier transformation results in the
current-current correlation function at the bosonic Matsubara
frequencies $i \omega_{n}$. The analytic continuation of this function 
onto real frequencies at zero momentum gives $\Lambda(0,\omega)$ of \eq{drudelinres}. 
However, the analytic continuation is a numerically ill-conditioned problem,
which we can avoid. 
As pointed out in section \ref{DRUDE},
one can take the limit $\omega\rightarrow 0$ directly along the imaginary axis. 
We therefore calculate $D$ by 
fitting $\Lambda$ on the Matsubara frequencies. 
This is still an ill-conditioned problem
when only information about a finite number of Matsubara frequencies
is available. It is therefore important to use a fitting function with
the proper analytical form.

{}From the analytic continuation of
\eq{ccmats} it can be seen that $\Lambda(0,i\omega_l)$ is a
well-behaved function on the imaginary axis,
namely a sum of Lorentz curves:
\beq{aprox}
  \Lambda(0,i \omega_{l}) \,=\, \sum_{j}\, c_{j} 
                                \frac{ \Delta_{j}}{\omega_{l}^{2} + \Delta_{j}^{2}}.
\eeq
We approximate it by a finite series
\beq{appro2}
    \Lambda_{\rm{FIT}}(0,i \omega_{l})\,=\, \frac{a}{\omega_{l}^{2}\,+\, b^{2}}
                                    \,+\, \frac{c}{\omega_{l}^{2}\,+\, d^{2}}
                                    \,+\, \frac{e}{\omega_{l}^{2}\,+\, f^{2}} ,
\eeq
and determine the constants $a,b,c,d,e,f$ by a $\chi^{2}$-fit. 
This ansatz fits the data very well. 
Fig.\ \ref{four} shows an example of our fit.
We find that the third term in \eq{appro2}, $e/(\omega_l^{2}+f^{2})$, 
and often also the second term,
do not contribute significantly, indicating that \eq{appro2} is a good approximation.
However, it should be noted that the constants $a,c$ and $e$ 
are not identical to the differences of the energy eigenvalues
of $H$ appearing in $\Lambda(0,i\omega_l)$,
since they result from a fit to a truncated series.
We have verified that our procedure to calculate the Drude weight
produces results compatible with those using the f-sum rule \eq{fsum}, 
but with smaller errors.

\begin{figure}[t]
\psfig{file=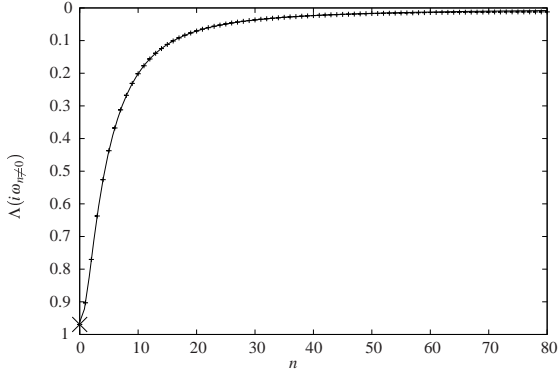,width=8cm}
\caption{\Label{four} Fit of $\Lambda(0,i\omega_n)$, at the nonzero Matsubara frequencies,
for $L=32$, $U=4$, $\mu=2.2$ and $\beta=10$, plotted {\em vs} $n$.
The cross marks the extrapolated value at $\omega=0$.
}\end{figure}

%
\begin{figure}[t]
 \psfig{file=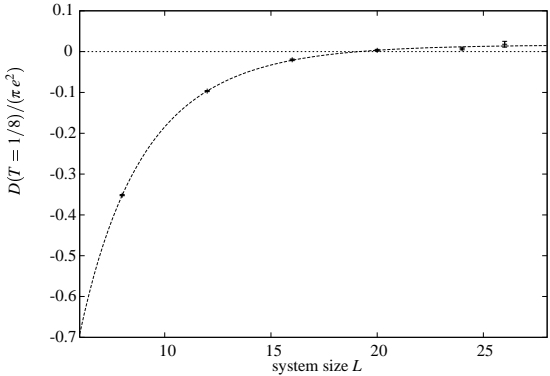,width=8 cm}
\caption[*]{\Label{five}
 The Drude weight at $\beta=8$, $U=4$, $\nu=1$
 is nonzero in the thermodynamic limit $L\rightarrow\infty$.
 The form of the interpolating function is motivated by the $T=0$ 
 Bethe ansatz results.
}\end{figure}
%

We used this procedure to determine $D(T)$ 
{}from finite temperature Quantum Monte Carlo runs.
We typically collected about $180 000$ Monte Carlo sweeps for each
data point, with a discrete time step of $\Delta\tau=1/32$ for
$\beta=3$ and $\Delta\tau=1/10$ for $\beta=8$.
We obtained the following results.

\Fig{five} shows the Drude weight 
for  repulsive interaction $U=4$ and inverse temperature 
$\beta=8$. 
The system size dependence is compatible to that of $D^{triplet}(L)$, \eq{Dspin}, as
expected for temperatures small compared to the charge gap.
\beq{DTbehavior}
  D(L)/(\pi e^2) \,=\, -A \cdot L^{-B} \exp[-C \cdot L] \,+\,d.
\eeq
We determined the parameters  by a $\chi^{2}$-fit.
For the case shown in fig.\ \ref{five} the result is
$A = 10.12  \pm 1.14,\; 
 B =  0.75  \pm 0.88,\;
 C =  0.22  \pm 0.09,$ and
$d =  0.020 \pm 0.003$.

Thus, in the thermodynamic limit, we obtain a finite Drude weight
$D(\beta=8,\nu=1,U=4)/(\pi e^2)=0.02\pm0.003$.
The system size behavior of the full Drude weight 
$D(L,\beta=8)$ (\eq{DTbehavior}) appears to be similar to 
that of the partial Drude weight $D^{triplet}(L,\beta)$
obtained by Bethe ansatz in \eq{Dspin},
within the large error of $B$,
even though $D(L,\beta=8$) contains charge exitations, as well as
higher spin and charge-spin-excitations.

Towards higher temperatures, the Drude weight increases rapidly.
Results at $\beta=3$ are shown in \fig{six}.
The approach to the asymptotic value is very fast now, and 
there is no longer a clear exponential behavior.
\begin{figure}[t]
\psfig{file=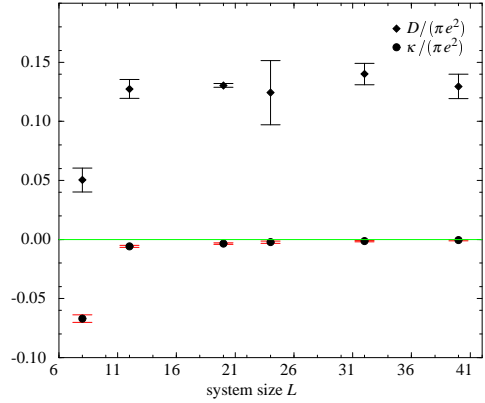,width=8 cm}
\caption{\Label{six} At large temperature ($\beta=3,U=4,\nu=1$)
the Drude weight is large and converges rapidly with system size.
The Meissner fraction $\kappa$ (see section \ref{MF}) converges to zero quickly.
}\end{figure}

We checked that for lattice sizes $L=4n+2$ the finite value of the Drude weight
remains the same within errors.
One might be concerned that at temperature  $\beta=3$ and lattice size $L=40$
the effects of the finite size gap could still be important.
We therefore verified that at very high temperature ($\beta=0.5, L=20$)
the Drude weight remains finite. In fact, its value increases further,
to about $D/(\pi e^2)=0.23(1)$.
In \fig{Abb.4.4a} our results for the Drude weight at various temperature in the half-filled 
case are shown.  
\begin{figure}
\psfig{file=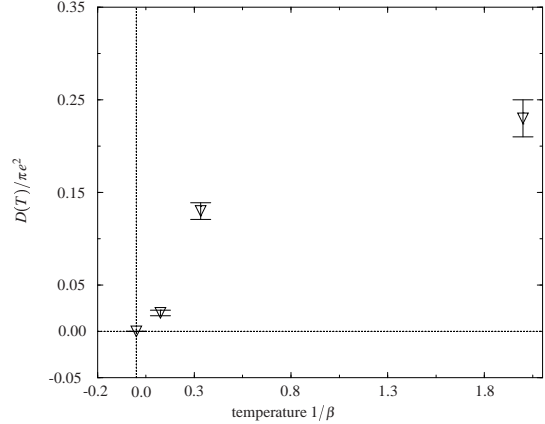,width=7.5 cm}
\caption{\Label{Abb.4.4a} The Drude weight in the repulsive model ($U=4$,$\nu=1$)
increases rapidly with temperature.
}\end{figure}

\begin{figure}[t]
\psfig{file=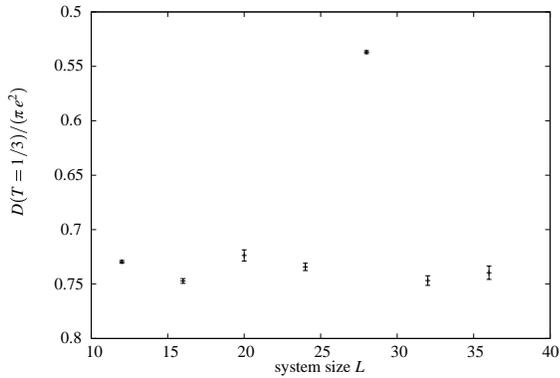,width=8 cm} 
\caption{\Label{Abb.4.4} At quarter filling $\nu \approx 0.500$ ($\beta=3$, $U=4$)
the Drude weight is finite.
Small variations in filling of the grand canonical simulations 
may be responsible for the large deviation of the data at $L=28$.
}\end{figure}
%
{\em Away from half-filling}, 
the finite temperature Drude weight remains finite.
Results at quarter filling are shown
in \fig{Abb.4.4}.
The asymptotic value is  already reached for very small system sizes.
Note that away from half-filling, there was some fluctuation of the
actual filling for different $L$ in the grand canonical QMC, which is 
not reflected in the error bars. This can cause the outlying point
at $L=28$ in \fig{Abb.4.4}.

\begin{figure}[t]
\psfig{file=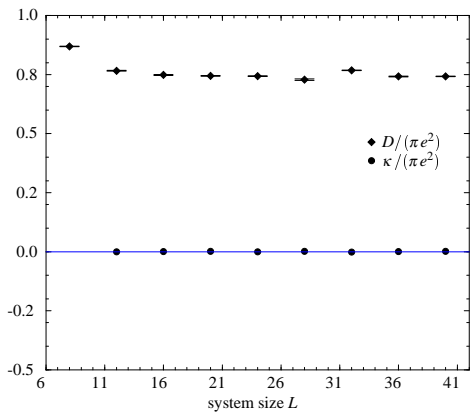,width=8 cm}
\caption{\Label{Sattr} Attractive Hubbard model at high temperature: 
$\beta=3,~U=-4$, $\nu=1$.
The  Drude weight (upper points) quickly converges to a large finite value,
whereas the Meissner fraction $\kappa$ vanishes (lower points; see section \ref{MF}).
For comparison:  $\VEV{k_{x}} \approx -0.890$, with little size dependence.
}\end{figure}
\begin{figure}[t]
\psfig{file=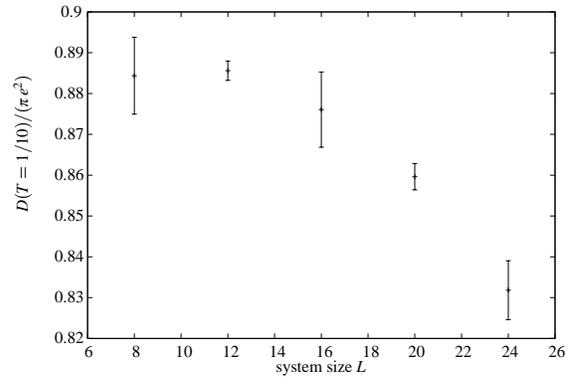,width=8 cm}
\caption{\Label{lowtemp} The Drude weight in the attractive model at
lower temperature:  $\beta=10,~U=-4,~\nu=1$
shows strong finite size dependence.
}\end{figure}
%
For the {\em attractive Hubbard model} at half-filling 
the situation is similar to the repulsive case.
At low temperatures the system size dependence is slower,
as shown in fig.\ \ref{lowtemp},
and the Drude weight has not yet reached its asymptotic value in our calculations.
At high temperature, shown in fig.\ \ref{Sattr},
the Drude weight converges quickly to a large value.

We see that in the half filled Hubbard model at finite temperature
the Drude weight is non-zero in the thermodynamic limit $L\rightarrow\infty$, 
both for the repulsive and for the attractive case.
Since half-filling is the insulating case of the Hubbard-model at zero 
temperature, this result is in disagreement with conjecture (2) (section \ref{INT}).

\section{\label{MF} The Meissner Fraction}
A  property closely related to the Drude weight is the 
Meissner fraction $\kappa$ \cite{Giamarchi-Shastry95} defined as 
\beq{kappa}
  \kappa \, \equiv \, \frac{\pi}{L} 
         \frac{\partial^{2} F}{\partial \Phi^{2}}\Big|_{\Phi=0, \vec{q}\rightarrow 0},
\eeq
where $F$ is the free energy.
It reduces to Kohn's Drude weight for $T=0$.

Evaluation of \eq{kappa} results in 
\beq{kappaalt}
  \frac{\kappa}{\pi e^2} \, \,=\,
    -\,\VEV{k_{x}}\,-\,\Lambda(\vec{q}\rightarrow 0,i \omega_{n}=0).
\eeq
Note that $\Lambda(\vec{q}\rightarrow 0,i \omega_{n}=0)$ at 
the zeroeth Matsubara  frequency, given in \eq{ccmat0},
is in general not equal to the analytically continued correlation function 
at $\omega\rightarrow 0$, \eq{limit},
so that at finite temperature $\kappa$ differs from the Drude weight $D$
by the contributions from degenerate states
\beq{degen}
 D(T) - \kappa = \lim_{\vec{q}\rightarrow 0}
  \frac{\beta}{L} \sum_{\stackrel{n,m}{E_{m} = E_{n}}} P_{n} 
                       \abs{\bra{n}j^p(\vec{q}) \ket{m}}^{2}.
\eeq
In the thermodynamic limit $L\rightarrow\infty$  and in a transverse
vector field, $\kappa$ measures the superfluid density 
\cite{Scalapino-White-Zhang93,Remark,Giamarchi-Shastry95}.
In our one-dimensional case it will thus become zero.
For a finite system, the Meissner fraction can be finite.

\begin{figure}[t]
 \psfig{file=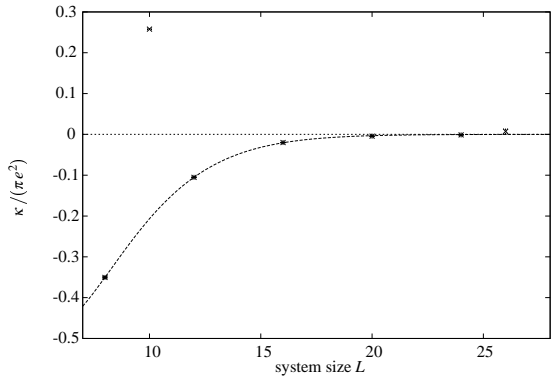,width=8 cm}
\caption{\Label{2S3}  
 Meissner fraction in the repulsive model at low temperature: $\beta=8$ and
 $U=4,~\VEV{k_{x}} \approx -0.955$, $\nu=1$. The two values at 
 $L=10$ and $L=26$ differ in sign, since they belong to system sizes where
 $L=4\cdot n +2$.   
}%
\vspace*{-0.3 cm}
\end{figure}
\begin{figure}[t]
\psfig{file=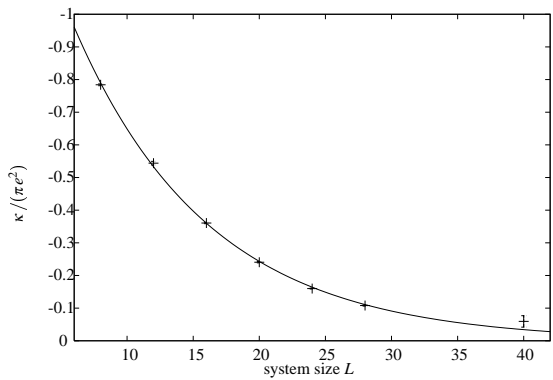,width=8 cm}
\caption{\Label{2Sneu} Meissner fraction in the attractive model at
  low temperature, $\beta=10$, $U=-4,~\nu=1$:
$\kappa \sim$ $ \exp[-(0.1 \pm 0.002) \cdot L]$.  
The kinetic energy is $~\VEV{k_{x}}\approx -0.96$ for all system sizes.
}\end{figure}
%

We show results for the Meissner fraction on finite systems \cite{Remark} 
in figures \ref{six},\ref{Sattr},\ref{2S3} and \ref{2Sneu}.
For $L\rightarrow\infty$, $\kappa$ vanishes.
We find a finite size behavior similar to that of the Drude weight.
As for the Drude weight at $T=0$ (see \fig{Abb2.1}), 
$\kappa$ is positive when $L=4\,n +2$ and negative when $L=4\,n$.
At high temperature $\beta=3$, $\kappa$ converges to zero very quickly 
both for the repulsive and for the attractive model 
(figures \ref{six} and \ref{Sattr}).
At low temperature the approach to zero is exponential in system size,
but much slower, as shown in figures \ref{2S3} and \ref{2Sneu}.
For very small systems, $\kappa$ is of similar magnitude as the Drude
weight.
Even though the difference $D-\kappa$, \eq{degen}, contains a factor of $\beta$,
this difference turns out to be {\it smaller} at larger $\beta$
in the temperature range studied,
both for small and for large systems, 
contrary to previous expectations \cite{Giamarchi-Shastry95}.

\section{\label{HEXT} Extended Hubbard model} 
In section \ref{QM} we showed that for the integrable Hubbard model \eq{hubb-model}
the finite temperature Drude weight both at half-filling and away from it 
is nonzero even in the thermodynamic limit. This is in contrast to the $T=0$ case,
where the Drude weight vanishes at half-filling due to the charge gap.

Let us now examine a nonintegrable case, provided by 
the extended Hubbard model with nearest neighbor interaction \cite{Montambaux-Poilblanc93}
\beq{extend}
  H_{e} \,=\, H_H \,+\, V\, \sum_{i} (n_{i \uparrow} + n_{i \downarrow}) (n_{i+1 \uparrow} 
         + n_{i+1\downarrow}).
\eeq

%
\begin{figure}[tb]
\psfig{file=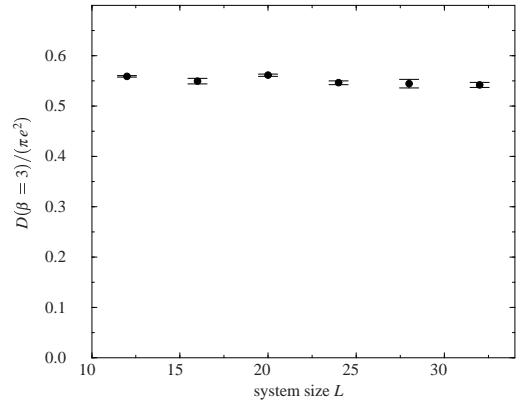,width=8 cm}
\caption{\Label{seven} 
The Drude weight in the extended Hubbard model at
at  $U=4$, $V=1/2$, $\beta=3$ and $\nu=0.8978 \pm 0.0002$ is nearly constant. 
The error bars do not take into account that the filling 
has some variation with $L$ in our grand canonical simulations.
}\end{figure}
%
We have measured the Drude weight and the Meissner fraction 
at finite temperature in this model.
Results for a weak extended interaction of  $V=1/2$ and away from half-filling
are shown in \fig{seven}.
The Meissner fraction drops to zero quickly, similar to the behavior 
in \fig{six}.
The Drude weight has a large value already at small system sizes.
It shows a very slow falloff, consistent with
$ D \sim  e^{-(0.00081 \pm 0.00004) L} $. 
This falloff may entirely be caused by a very small 
systematic increase in the filling factor 
measured in the grand canonical simulations.
If the exponential falloff is real, then the Drude weight would drop
to small values only at extremely large system sizes,
but would be zero in the thermodynamic limit,
in agreement with the conjecture of Zotos et al..
However,
{}from our data it appears more likely that the
thermodynamic limit of $D$ is finite, which would be in disagreement with
conjecture (3). This would imply that, even though the model 
is not integrable, there is an operator satisfying \eq{schwarzrule}
in the extended Hubbard model.
A similar conclusion has been drawn for a related model recently \cite{Prosen98}.

We have also measured the Drude weight at $\beta=5, U=4$ and larger
repulsion $V=1.6$, on a large system $L=60$.
Both at  quarter filling $\nu=0.5$ and at  $\nu=0.633$ 
we find a finite value $D/(\pi e^2)=0.21(2)$,
even though the model is not integrable.

\section{Conclusions}
%
We have presented results for the Drude weight $D(T)$,
on mesoscopic systems and extrapolated to the thermodynamic limit,
for zero and finite temperature in several versions of the 1D
Hubbard model,
including the dependence on system size, filling, and temperature.

At zero temperature we provided the full filling dependence of $D$,
by solving the Bethe ansatz equations on large systems.
For small finite temperatures we computed the dominant contribution
to $D(T,L)$ on small systems from the Bethe ansatz.
At larger finite temperature we used Quantum Monte Carlo.
We showed that the Drude weight can be obtained by an extrapolation
of the current-current correlation function purely in imaginary frequencies.
We found a non-vanishing Drude weight at finite temperature for all
cases considered,
the repulsive  Hubbard model both at and away from
half-filling, the attractive model at half filling,
and the extended Hubbard model away from half filling.
The Drude weight quickly grows with temperature for the half
filled Hubbard model.

Our results for the
integrable half-filled Hubbard model do not confirm 
conjecture (2) (section \ref{INT}) 
on the connection between integrability and finite temperature Drude weight,
and we find that conjecture (3) would disagree with
the (likely) non-zero value of the Drude weight
in the nonintegrable extended Hubbard-model.

\begin{acknowledgements}
We thank B. Brendel, G. Bed\"{u}rftig, A. Sandvik, and X. Zotos for useful 
and enlightening discussions,
and C. Gr\"ober and M. Zacher for support with the calculations.
One of us (W. Hanke) acknowledges many illuminating interactions 
with D.J. Scalapino.
This work was supported by the BMBF (05 605 WWA 6 and XF05 SB8WWA 1).
The computations were performed at the HLRZ J\"{u}lich and the HLRS Stuttgart.
\end{acknowledgements}

\begin{appendix}
\section{Spectral decomposition of the Drude weight}
The Drude weight is given by 
\beq{drudelinresa}
   \frac{D(T)}{\pi e^2} \,=\, -\VEV{k_{x}} \,-\, \Re\{ \Lambda(0, \omega \rightarrow 0)\}
\eeq
at $\Phi =0$. Here $\VEV{k_{x}}$ is the average kinetic energy per site and
$\Lambda(\vec{q} ,\omega)$ is the current-current correlation
function in frequency space \cite{Scalapino-White-Zhang93}.
It can be obtained from
analytic continuation of
\beq{correlation}
  \Lambda(\vec{q} ,i \omega_{n})\,=\, \frac{1}{L} \int_{0}^{\beta} e^{i \omega_{n} \tau}
 \VEV{j^{p}(\vec{q},\tau)j^{p}(-\vec{q},0)}\,d\tau
\eeq
where $\beta=1/T$, $i\omega_{n}=2\pi i n/\beta $  are the Matsubara frequencies,
$\vec{q}$ is the momentum of the applied external vector potential,
and $e\,j^{p}(\vec{q})$ is the Fourier transform of the paramagnetic current density,
\begin{equation}
  \label{current-density}
  j^{p}(\vec{q})\,=\, i\,t\sum_{\scriptsize
       \begin{array}{l}{k=1,..,L}\\
       {\sigma=\uparrow,\downarrow}\end{array}}\, 
      e^{-i \vec{q} \vec{x}_k}\,
 (c^{\dagger}_{k+1,\sigma}c_{k,\sigma}\,-\,c^{\dagger}_{k,\sigma}c_{k+1,\sigma}) \,.
\eeq

In an eigenbasis of the Hamiltonian \eq{hubb-model}, the
current-current correlation function at nonzero frequency is given by
\beq{ccmats}
    \Lambda(\vec{q},i \omega_{l} \neq 0) =\,  
      \hskip-2ex \sum_{\scriptsize\begin{array}{c}{n,m}\\{E_n \ne E_m}\end{array}}\hskip-2ex
      \frac{P_n}{L} \frac{e^{( E_n - E_m) \beta} - 1}
                 { i\omega_l + E_n - E_m}               
            \abs{\bra{n}j^p(\vec{q}) \ket{m}}^{2},
\eeq
and exactly at zero Matsubara frequency it is
\beq{ccmat0} \begin{array}{lllcl}
  \Lambda(\vec{q},i \omega_{l} \!=\! 0) 
    &=& \frac{2}    {L} \bigsum{\mystack{n,m}{E_n \ne E_m}}
          &\frac{P_n}{ E_m -E_n}   &  \abs{\bra{n}j^p(\vec{q}) \ket{m}}^{2}   \\[7ex]
    &+& \frac{\beta}{L} \bigsum{\mystack{n,m}{E_n  =  E_m}}    
          &      P_n               &  \abs{\bra{n}j^p(\vec{q}) \ket{m}}^{2}.
\end{array}\eeq
Here $\VEV{n|j^p|m}$ are matrix elements of the current operator 
in the eigenbasis and $P_{n}=\exp(-\beta\,E_{n})$ denotes the Boltzman factor
for the $n$th eigenvalue of the Hamiltonian.

The current-current correlation function is well defined \cite{Baym} in the upper
complex plane 
by specifying its values on an infinite set of finite Matsubara frequencies, \eq{ccmats}.
The analytic continuation can easily be performed by taking 
$i\omega_l \rightarrow \omega+i\delta$ in \eq{ccmats}.
Performing the zero frequency limit $\omega \rightarrow 0$ then yields 
\beq{limit}
    \Re\left(\Lambda(\vec{q}, \omega \rightarrow 0)\right) \,=\,  
        \frac{2}{L} \sum_{\stackrel{n,m}{E_{m} \neq E_{n}}} \frac{P_{n}}{ E_{m}-E_{n}} \, 
       \abs{\bra{n}j^p(\vec{q}) \ket{m}}^{2},
\eeq
which is identical to the first term in \eq{ccmat0}.
Thus
\beq{Drude1}
   \frac{D(T)}{\pi e^2} \,=\, -\VEV{k_{x}} \,-\, 
        \frac{2}{L} \sum_{\stackrel{n,m}{E_{m} \neq E_{n}}} \frac{P_{n}}{ E_{m}-E_{n}} \, 
       \abs{\bra{n}j^p(0) \ket{m}}^{2}.
\eeq
Using second order perturbation theory
and assuming that $\Phi$ removes all degeneracies, 
the  expression  takes  the same form as \eq{Kohn}, so that 
then the two definitions \eq{ext} and \eq{ext2} agree.\\
The second term in \eq{ccmat0} is a nonanalytical part of the
thermal Greens function \cite{Kwok}, which does not contribute to the Drude weight,
but instead to the Meissner fraction, discussed in section \ref{MF}.
\end{appendix}



\begin{references}
\bibitem{Remark} 
  The spatially constant $\phi$ on the ring employed here, as is customary,
  corresponds to a finite momentum $\vec{q}$ for the electromagnetic vector potential 
  in the three dimensional real space in which the ring is embedded.
  The momentum vanishes in the thermodynamic limit.
\end{references}
\end{document}